\documentclass[journal]{IEEEtran}
\usepackage{graphicx}
\usepackage[fleqn]{amsmath}
\usepackage{cite}
\usepackage{amssymb,epsfig}
\graphicspath{{images/}}
\usepackage{epstopdf}
\usepackage{placeins}

\ifCLASSINFOpdf
\else
\fi
\begin{document}
%
\title{Active Decoupling of Transmit and Receive Coils for Full-Duplex MRI}
%
%
%

\author{Maryam~Salim,~\IEEEmembership{Member,~IEEE},
        Ali Caglar Ozen, Michael Bock, Ergin Atalar}

\maketitle

\begin{abstract}
Objective: Concurrent excitation and acquisition in MRI is a method to acquire MRI signal from tissues with very short transverse relaxation time. Since transmit power is many orders of magnitude larger than receive signal, a weak coupling dominates the MR signal during CEA. Thus, appropriate decoupling between transmit and receive coils is required. In this study, two controllable decoupling designs are investigated for achieving isolation between coils.
Methods: A modified version of isolation concept used in the full-duplex radios in communication systems is applied to acquire MRI signal using CEA. In our new method, a small copy of RF transmit signal is attenuated and delayed to generate the same coupling signal which is available in the receiver coil. Then it is subtracted from the receive signal to detect the MRI signal. The proposed decoupling method is developed and implemented in two designs: Semi-Automatic and Fully-Automatic Controllable Decoupling Designs.
Results: Using Semi-Automatic Controllable Decoupling Design, decoupling of more than 75~dB is achieved. Fully-Automatic Controllable Decoupling Design provides more than 100~dB decoupling between coils which is good enough for detecting MRI signals during excitation from tissues with very short transverse relaxation time.
Conclusion: This study shows feasibility of applying full duplex electronics to decouple transmit and receive coils for CEA in a clinical MRI system.
Significance: These designs can automatically tune the cancellation circuit and it is a potential tool for recovering signal from tissues with very short T2 in clinical MR systems with a minor hardware modification.
\end{abstract}

\begin{IEEEkeywords}
MRI, CEA, Decoupling, Full-Duplex.
\end{IEEEkeywords}

%
\IEEEpeerreviewmaketitle

\section{Introduction}
\IEEEPARstart{M}{agnetic} Resonance Imaging (MRI) is a powerful diagnostic imaging tool which shows soft tissues like muscles and fats with high contrast. However, MRI is rather poor in imaging tissues like bones and lungs~\cite{short-t2}. The MR signal from cortical bone, tendons, ligaments and menisci mostly contain short transverse relaxation time, $T_2$, components. Only their long $T_2$ components are visible in conventional MRI scanners. In conventional MRI techniques transmit mode and receive mode are separated by a time delay which is typically too long to allow detection of the materials with very short transverse relaxation time, ($T_2$)~\cite{ultrashort-te,ImagingLungs}. Therefore, there is no signal or a little signal available in the receiver coil.

One approach to image the short $T_2$  components of these tissue is zero time echo (ZTE). In ZTE, the dead time (or acquisition delay) between the end of RF excitation and the start of data acquisition sets a limit to the lowest achievable echo time (TE) or results in missing data points at the center of k-space~\cite{MRI-ZTE-2012}. Also the acquisition dead time because of RF pulsing, transmit-receive switching (T/R), signal filtering time, and analog to digital conversion time. Therefore, this dead time puts limits on the lowest achievable echo time~\cite{MRI-ZTE-2011,NMR-Data}. Another approach for getting signals from these tissues is known as concurrent excitation and acquisition (CEA). In this method there is not any dead time due to transmit/receive switching mode and both transmit and receive coils are operated simultaneously. The main difficulty in CEA is coupling between transmit and receive coils. MR signal is very weak and because transmit power is many orders of magnitude larger than receive signal, even a weak coupling might dominate the receive signal and MR signal becomes not detectable. The performance of the CEA imaging technique depends on the amount of isolation. Ideally the coupled signal should be less than the noise floor. By using the CEA technique one can get the MR signal continuously even during the RF excitation pulse. An advantages of this method is that it uses very low RF peak power. Also, in comparison with the previous methods, it has no dead time and so enables getting images from the tissues with very fast decay time. In this method the peak of RF power is reduced by at least an order of magnitude~\cite{Continuous-wave-MR,Development-3-D}.

Several studies demonstrated MRI with CEA such as continuous SWIFT~\cite{SWIFT}, and rapid scan correlation spectroscopy~\cite{Rapid-ScanNMR}. In SWIFT (SWeep Imaging with Fourier Transformation), a hybrid coupler system is connected to a dual coil system and the NMR signal is acquired during continuous radio frequency excitation. The hybrid coupler has a phase difference of 180 degrees between the RF input port and the output port to the receiver coil. Therefore, it subtracts RF signal which is available in the output signal from input RF transmit signal for getting the MR signal. However, using this circuit needs accurate tuning of the isolator and it is sensitive to small changes in the impedance of coils. Maximum achieved decoupling value in the literature for CEA imaging is 60 dB [cite] which needs to be improved further to get closer to the noise floor. Therefore, the rest of the work is carried out after data acquisition by using cross-correlation method and MR signal should be extracted from the acquired data~\cite{SWIFT,SWIFT-ISMRM}.

MRI with concurrent RF excitation and signal acquisition is promising approach to eliminate the acquisition delays with true zero echo times. For isolating the RF surface coil there are different decoupling techniques such as partial overlap of coils~\cite{NmrPhaseArray}, capacitive and inductive decoupling~\cite{Ladder-networks,SingleCouplingSurfaceCoil,8-channel,LC-Decoupling}, geometrical decoupling~\cite{nuclear}, and active decoupling \cite{caglar}. In overlapping method, adjacent coils are placed overlap to make the mutual inductance as low as possible and also the coils are connected to low input impedance pre amplifiers to reduce the coupling between the coils which are not overlapped~\cite{NmrPhaseArray}. In capacitive and inductive method, either capacitors or inductors are placed between the two coils so that the mutual inductance of the two coils is cancelled. The inserted capacitor methods is also extended to decouple the nearest and the non-nearest neighbour coils by using a capacitor network~\cite{Ladder-networks,SingleCouplingSurfaceCoil,8-channel}. In geometrical decoupling, placing two individual coils as transmit and receive coils in certain position and orientation, can reduce the RF signal which is transmitted to the receive coil due to the coupling. In this method most of the RF excitation pulse in the receive coil will be vanished. However, the amount of isolation achieved by this method is limited~\cite{nuclear}. Prior to this work, \cite{caglar,caglar-ISMRM} has proposed an active decoupling method which uses an extra transmit (Tx) chain to generate a cancellation signal that is combined with the signal on the receive (Rx) chain. However, using two different Tx chains with different random noises causes the noise in the system to increase and hence SNR to decrease. The best reported isolation achieved using this design is 70~dB.

A similar problem exists in communication systems, in order to transmit and receive in the same channel simultaneously (full-duplex radios), transmit and receive signals should be isolated. To achieve full duplex, a radio has to completely cancel the significant self-interference that results from its own transmission to the received signal. If self-interference is not completely cancelled, any residual self-interference acts as noise to the received signal and reduces SNR and consequently throughputs~\cite{full-duplex}. In this work, we have used the concepts used in the full-duplex radio system to suppress the Tx signal coupled on the Rx coil. The key insight of this work is that in fact we know the signal we transmit and we are only designing an analog circuit to subtract the transmit RF signal from the receiver coil. Our technique is implemented using an unequal Wilkinson power divider~\cite{unequal} to create a small copy of the transmit signal and adjusted its delay time and attenuation amount using phase shifters and programmable attenuators~\cite{att} to cancel the coupling signal. Our design also uses two coils which are geometrically decoupled from each other, one for transmit purpose and the other one for receiver. We could achieve up to 100~dB decoupling between Tx and Rx coils using our cancellation circuit which is much closer to the noise floor. This method can also cancel input random noise. The conductive sample phantom was then placed at the imaging volume of the setup. We implemented the design and optimize it to produce the best performance. We demonstrated feasibility of MRI with CEA using our decoupling method which is potentially useful for MRI of tissues with very fast decay time.

\section{Material and Method}
Our design is based on two kinds of decoupling; geometrical decoupling and active decoupling. Geometrical decoupling is a conventional method to reduce the coupling of transmit and receive coils. The interaction between the electric and magnetic fields of transmit and receive coils is the origin of the coupling between them. In geometrical decoupling, we reduce the coupling between the coils by placing the coils in certain position and orientation in which the interaction between the electric and magnetic fields is minimum. By orienting the coils orthogonally and placing the center of the coils on a horizontal plane, we can reduce the RF signal which is transmitted to the receive coil due to the coupling. Ideally, the amount of magnetic flux in the receive coil should be zero in this case. However, since the polarization of two coils are not totally linear and it is slightly elliptical, the fields cannot cancel each other totally~\cite{Comparison-Polarization}. In other words, geometrical decoupling can remove the amount of mutual inductance between two coils and it cannot change the mutual resistance. Therefore, for achieving better decoupling active decoupling is necessary.

\begin{figure}[!ht]
	\centering
	\includegraphics[width=8.5cm]{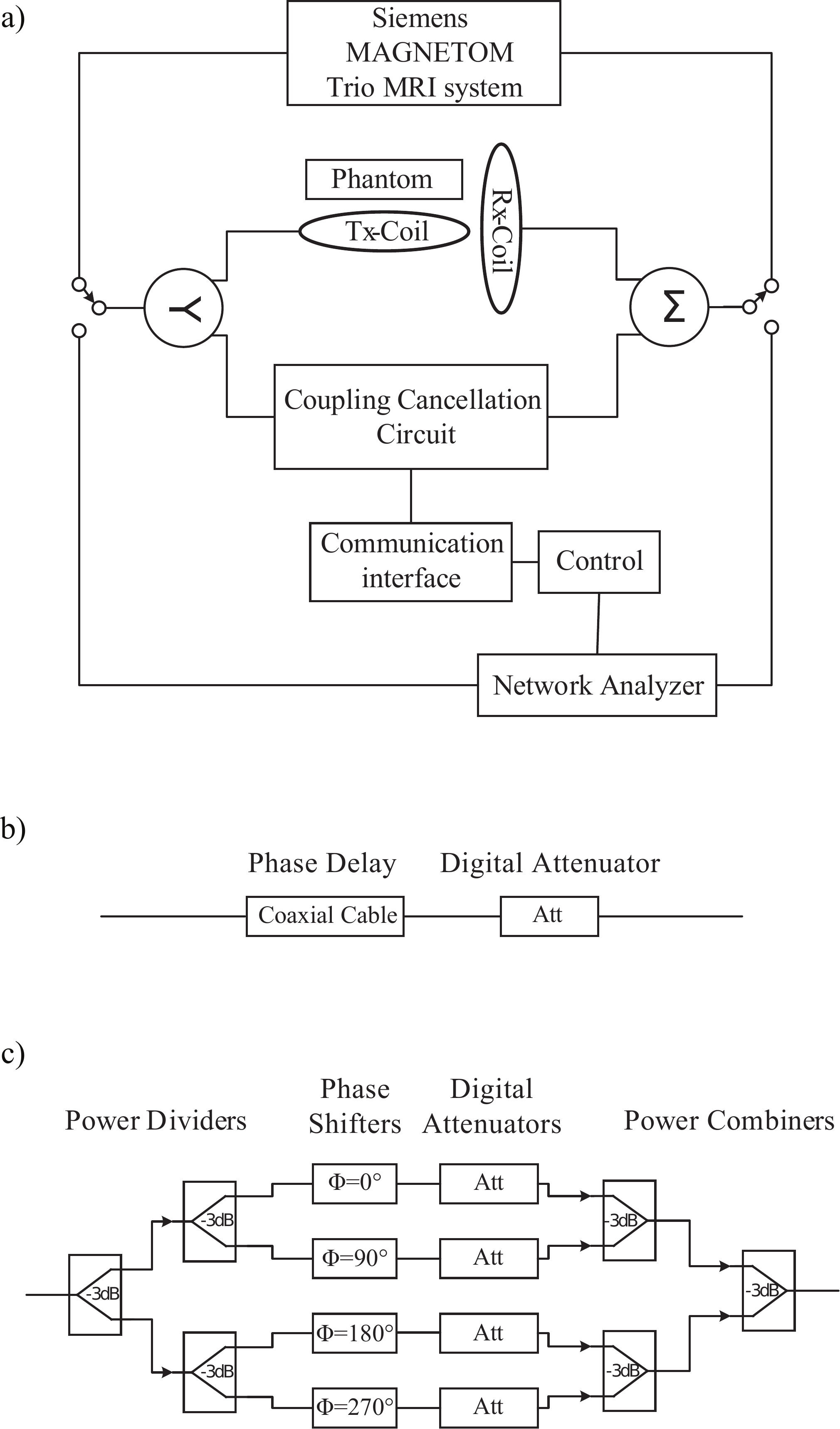}
	\caption{a) Schematic of the proposed decoupling method. The input RF signal is divided between the transmit coil and the coupling cancellation circuit. Then the output signals are added to get the MRI signal. Using a network analyzer the amount of decoupling is measured and the coupling cancellation circuit is optimized. b) Schematic of the semi-automatic controllable decoupling design. This design contains a coaxial cable for generating phase delay and a digitally controlled attenuator IC. c) Schematic of the fully-automatic controllable decoupling design containing four delay and attenuator lines.}
	\label{fig:method}
\end{figure}

In our experiments we fixed the geometric decoupling in order to provide approximately 25~dB isolation. This isolation is achieved by placing transmit and receive coils as described. As we discussed, the performance of the CEA imaging increases as the isolation between the coils increases. Therefore, active decoupling is required for imaging. Basically, the received RF signal is the delayed and attenuated version of the input RF signal due to the geometrical decoupling. Our analog cancellation circuit, which is a modified version of cancellation circuit used version of communication full-duplex radios, gets a sample of the input RF signal and applies the same attenuation and delay to the input RF signal and creates a copy of the received RF signal and then subtracts it from the received signal. Hence, the received RF signal will vanish from the received signal. Fig.~\ref{fig:method}(a) shows an overview of our proposed method for decoupling. Input RF signal is divided between the transmit coil and cancellation circuit using an unequal Wilkinson power divider~\cite{unequal}. Most of the signal reaches the transmit coil for increasing the power efficiency. Using a coupling cancellation circuit, the desired signal is generated and added back to the receive signal using a power combiner to properly isolate the coils. We used an E5061B Agilent network analyzer for measuring the amplitude of $ S_{12} $ of the output signal as a feedback for our algorithm and minimized that with tuning of the attenuation factor. Then, the setup was connected to the MRI system. The coupling cancellation circuit is implemented using two different designs.

\subsection{Semi-Automatic Controllable Decoupling Design}
\begin{figure}[!ht]
	\centering
	\includegraphics[width=7.5cm]{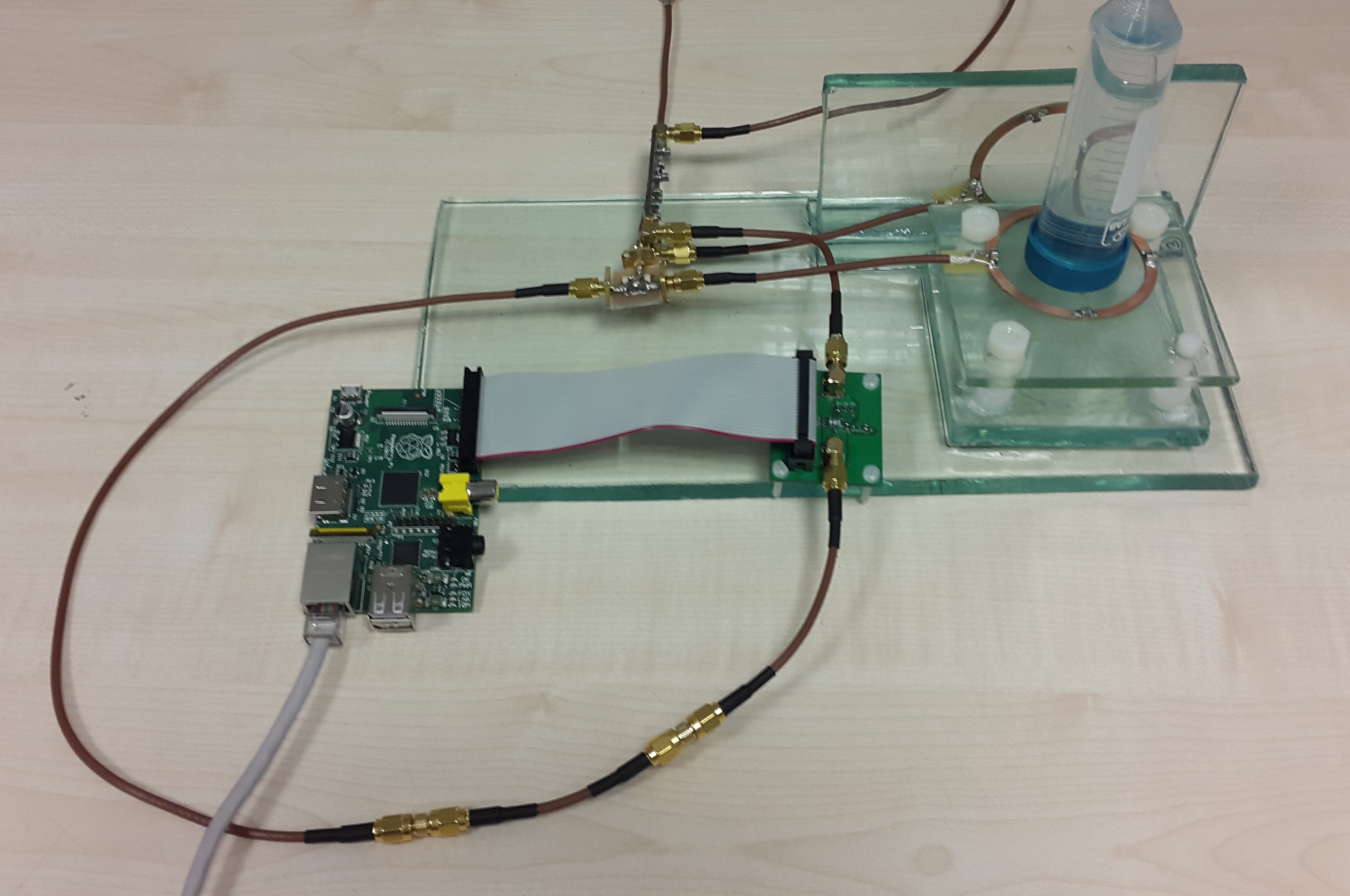}
	\caption{Image of the semi-automatic controllable decoupling circuit connected to the geometric decoupling setup.}
	\label{fig:1-line}
\end{figure}

Firstly, in order to prove our concept, we designed a cancellation circuit consisting a fixed delay line and a tunable attenuator. A small copy of the input RF signal is delayed and programmatically attenuated. The signal which is available at the output of the attenuator is then subtracted from the signal on the receive path. A tuning algorithm is used to find the best attenuation factor such that the RF signal coupled on the receive coil is minimum.
The fixed delay line is implemented using coaxial cables. We should pick the fixed delay in our cancellation circuit to be 180~degrees more than the delay of the coupling signal which we have in the receive coil~\cite{marconf}. For accomplishing this purpose, we measured the phase of $ S_{12} $ for two orthogonal coils. Then we prepared the suitable coaxial cable for achieving this phase delay.
A programmable attenuator (HMC759LP3E)~\cite{att} is used in the analog cancellation circuit (Fig.~\ref{fig:method}(b)). This attenuator can be programmed in steps of 0.25~dB from 0 to 31.75~dB for a total of 128 different values. The attenuated signal is then added with the signal which is achieved from the receiver coil using a Wilkinson power combiner~\cite{equal}. Fig.~\ref{fig:1-line} shows an image of the semi-automatic controllable decoupling setup.

\subsection{Fully-Automatic Controllable Decoupling Design}
In the second design, we used multiple delay and attenuator lines with fixed phase delays to achieve fully-automatic control. According to trigonometric identities, we can implement a variable phase delay using multiple fixed phase delays. We used four fixed delays which are 0, 90, 180, and 270 degrees, and attenuators to implement every phase delay and attenuation needed for duplicating the received RF signal. We have used three lumped-element phase shifters to generate 90, 180, and 270 degrees phase delay.

Fig.~\ref{fig:method}(c) illustrates the concept of our fully-automatic controllable decoupling method. Assuming that the input RF signal is single frequency, the input RF signal $Asin(wt)$ is divided and fed to the transmit coil and the cancellation circuit. The coils apply an attenuation and a phase delay to the input signal due to the geometric decoupling and we get  $A\alpha sin(wt+\phi)$ at the receive coil.~$\alpha$ is the coefficient of the total attenuations that input RF signal experiences to reach the receive coil which is due to the unequal power divider and the geometric decoupling. In order to suppress this signal, we have to generate $-A\alpha sin(wt+\phi)$ using the cancellation circuit. However, the input RF signal is a sinc function with a finite bandwidth and is not a single frequency signal. The proposed theory is applicable if we assume that A is not time-varying. This assumption makes our design very simpler but causes the decoupling results to be narrow-band with finite bandwidth.

In the analog cancellation circuit, we take the attenuation factor of the signal which is attenuated and delayed in each line and reached to the output power combiner without the effect of the programmable attenuators as. According to the Eqn.~\ref{eq:4-line}, we can produce $-A\alpha sin(wt+\phi)$ using two active lines:

\begin{equation}
\begin{aligned}
&-A\alpha sin(wt+\phi)\\
\label{eq:4-line}
	&=-A\alpha~(cos(\phi)sin(wt)+sin(\phi)sin(wt-\dfrac{\pi}{2}))\\
	&=Aa\beta~sin(wt+\phi_1)+Ab \beta~sin(wt+\phi_2)
\end{aligned}
\end{equation}

where \textit{a} and \textit{b} are the attenuation factors to be set to the programmable attenuators of the two active lines. $ \phi_1$ is either 0 or 180 degrees showing line~1 or line~3 as the first active line and $ \phi_2 $ is either 90 or 270 degrees showing line~2 or line~4 as the second active line. Actually since negative attenuation is not possible, two of four lines are selected to achieve the desired sign. Table~\ref{tb:phi} shows how the system selects two lines according to the phase delay of the coils ($ \phi $).

\begin{table}[!ht]
\caption{Number of two selected lines according to the phase delay between transmit and receive coils.}
\label{tb:phi}
\centering
\begin{tabular}{ c | c | }
\cline{2-2}
& Active Line Numbers 	\\\hline
\multicolumn{1}{ |c|  }{$ 0<\phi<90	$}		&	$ 3, 4 $		\\\hline
\multicolumn{1}{ |c|  }{$ 90<\phi<180 $	}	&	$ 1, 3 $		\\\hline
\multicolumn{1}{ |c|  }{$ 180<\phi<270 $}	&	$ 1, 2 $		\\\hline
\multicolumn{1}{ |c|  }{$ 270<\phi<360 $}	&	$ 2, 3 $		\\\hline
\end{tabular}
\end{table}

\begin{figure}[!ht]
	\centering
	\includegraphics[width=7.5cm]{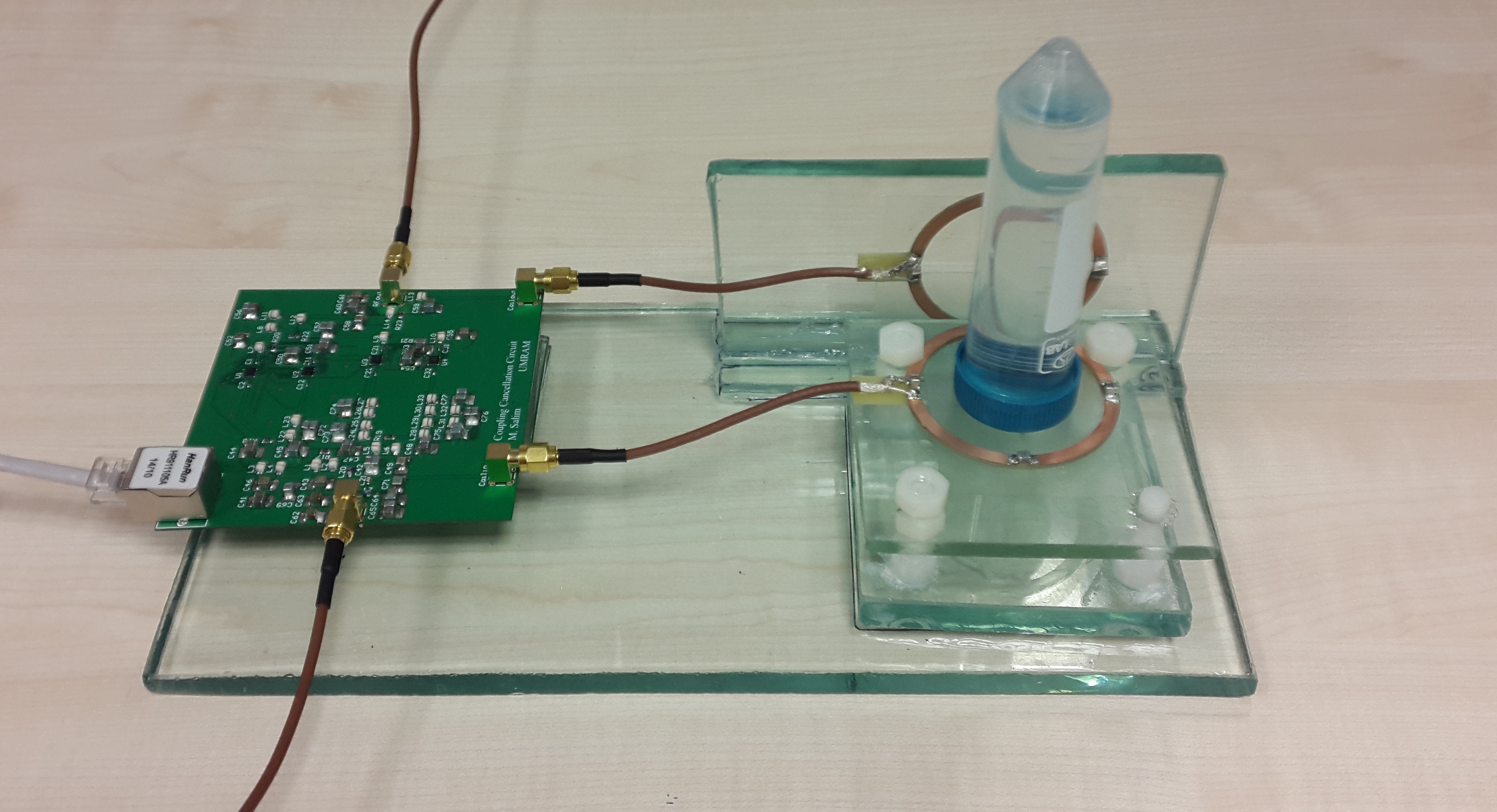}
	\caption{Image of the fully-automatic controllable decoupling circuit connected to the geometric decoupling setup.}
	\label{fig:4-line}
\end{figure}

It is clear that the attenuation factors \textit{a} and \textit{b} should be positive numbers less than 1. So, $ \alpha $ should be less than $ \beta $. In other words, the total attenuation that the RF signal faces to reach the receive coil should be greater than the attenuation that the RF signal faces in each line to reach the output power combiner without considering the effect of the programmable attenuators. In order to satisfy this condition as well as maintaining the efficiency of the setup by feeding the most of the power to the coils, we used an unequal power divider with ratio of 9 to 1. It means that the transmit coil is fed by 90~\% of the power and 10~\% of the power is passing through the cancellation circuit. We have used unequal lumped-elements Wilkinson configuration for power divider with reference characteristic impedance at all ports equal to 50~ohm~\cite{unequal}. Fig.~\ref{fig:4-line} shows an image of the fully-automatic controllable decoupling setup~\cite{martez}.

\section{Results}

\begin{figure}[!ht]
	\centering
	\includegraphics[width=7.5cm]{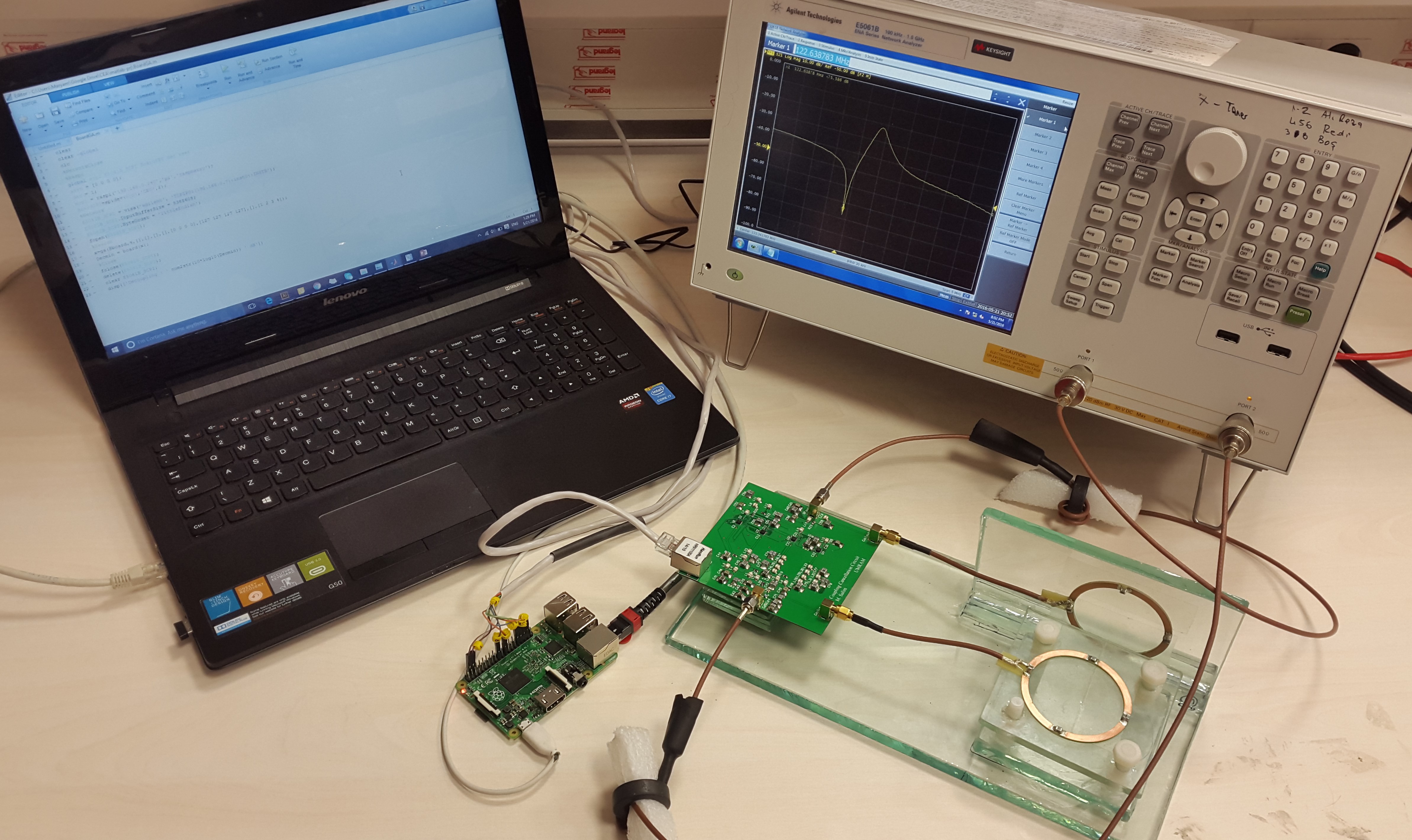}
	\caption{Experimental setup of fully-automatic controllable decoupling design.}
	\label{fig:setup}
\end{figure}

We did our experiments in a Siemens Magnetom 3~Tesla clinical MRI system with an eight channel parallel transmit array unit. Firstly, we put the setup into the MR magnet and connected the circuit to the control PC using a Raspberry Pi as the communication interface. The network analyzer was used to calculate the decoupling of the setup which was then fed to the genetic algorithm based optimization software written in MATLAB to find the best attenuation factors of all the lines and achieve the highest decoupling value (Fig.~\ref{fig:setup}). Then, the setup was connected to the MRI system.

\begin{figure}[!ht]
	\centering
	\includegraphics[width=8.5cm]{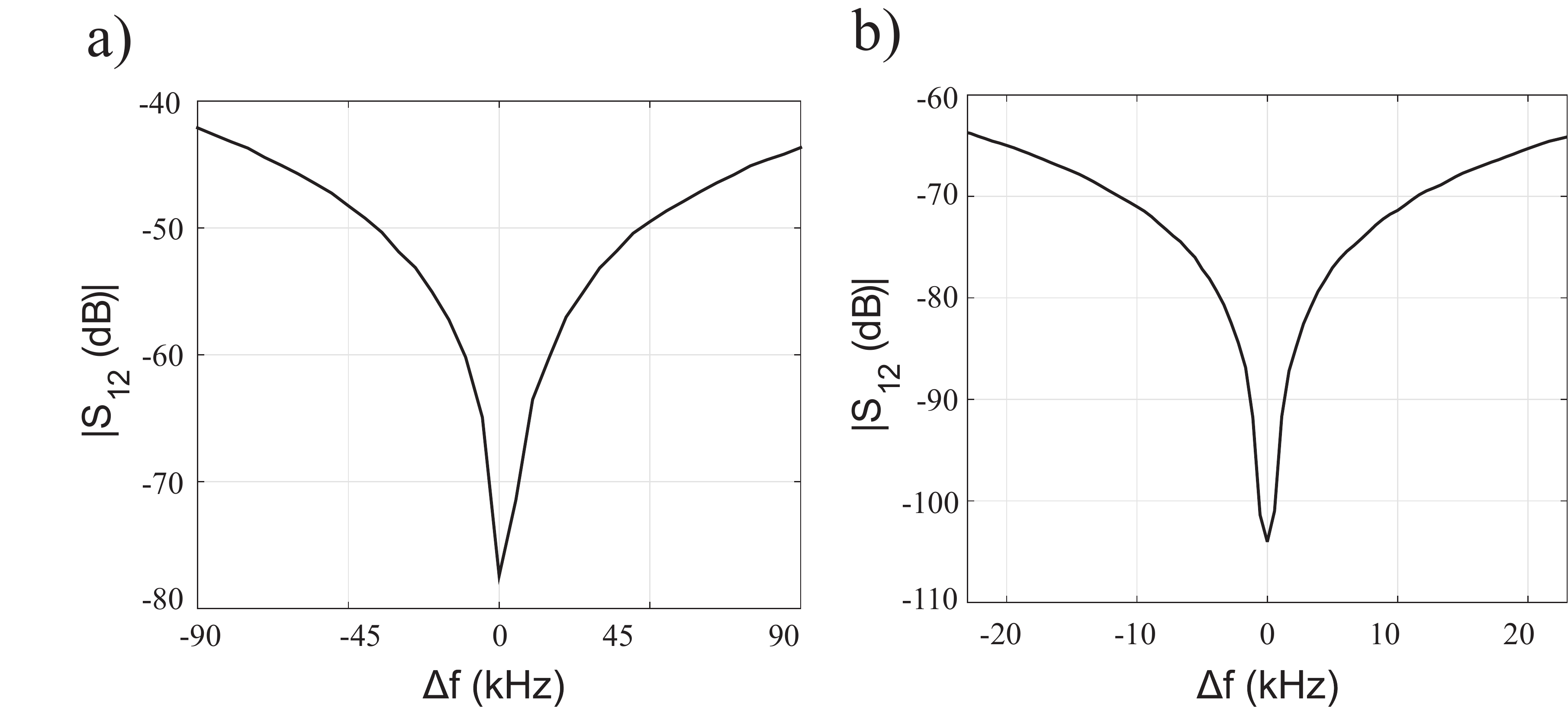}
	\caption{Amount of decoupling using a) semi-automatic and b) fully-automatic controllable decoupling design.}
	\label{fig:decoupling}
\end{figure}

Fig.~\ref{fig:decoupling} shows the decoupling amount achieved inside the MRI magnet using both decoupling designs. In semi-automatic design we achieved more than 75~dB of decoupling. Using fully-automatic design we could get 23~kHz bandwidth for 70~dB decoupling and we achieved more than 100~dB decoupling. In the fully-controllable design, the optimization algorithm digitally controls the circuit and ensures to have the best achievable decoupling as we change the phantom inside the setup.

\begin{figure}[!ht]
	\centering
	\includegraphics[width=8.3cm]{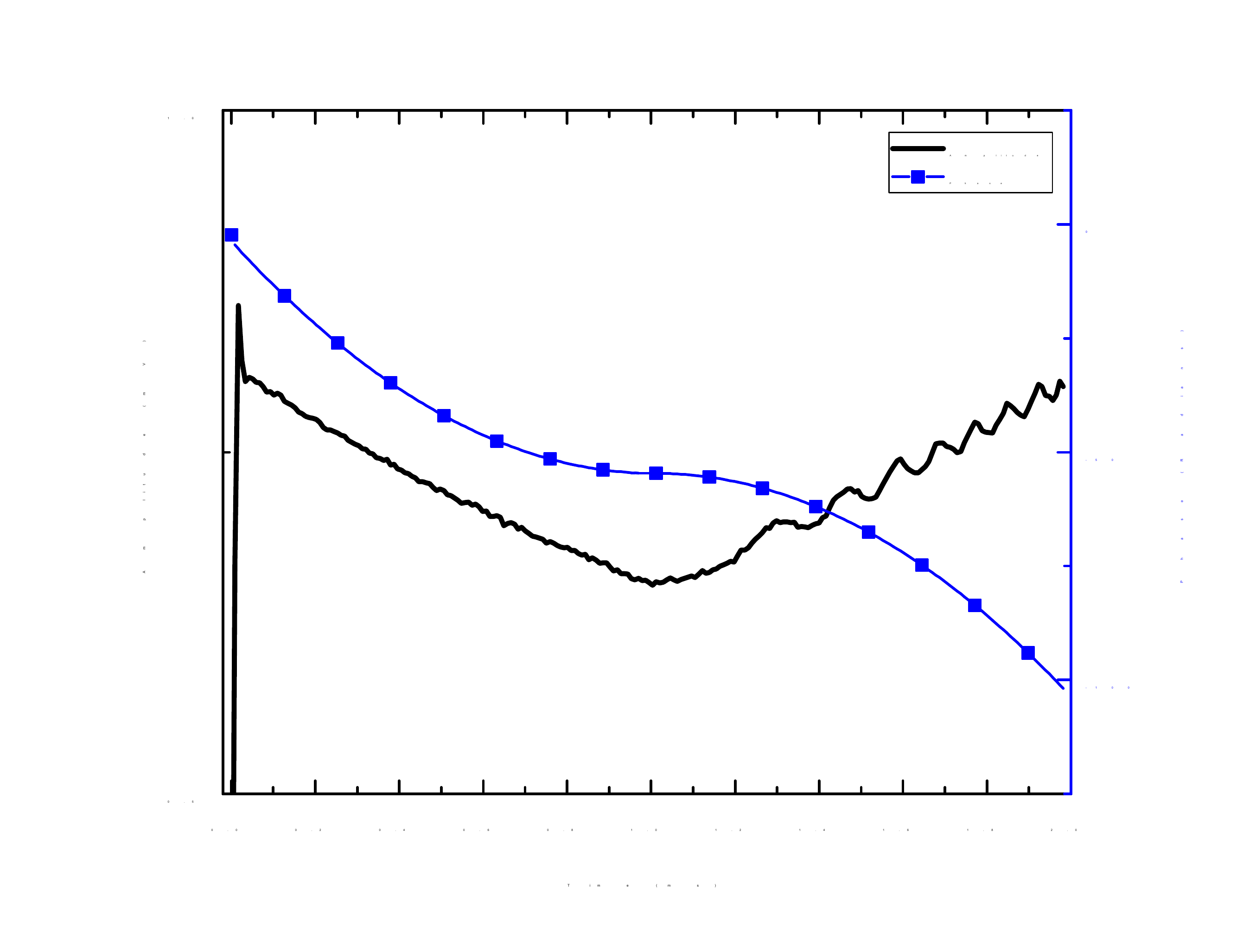}
	\caption{Phase and amplitude of the unprocessed CEA MR signal from a CuSO4(aq) sample. Spins get excited as soon as the resonance conditions are satisfied during the linear frequency sweep.}
	\label{fig:MR-Signal-PA}
\end{figure}

Fig.~\ref{fig:MR-Signal-PA} shows the raw-data for CEA with a CuSO4(aq) phantom using semi-automatic controllable decoupling design. The oscillations due to the MR signal response are already visible without any data processing. Note that the first 4 points of the each acquired radial spoke were neglected in reconstruction since the data was deformed by the ADC filtering effects. Missing points were interpolated using the consecutive points.

\begin{figure}
	\centering
	\includegraphics[width=7.5cm]{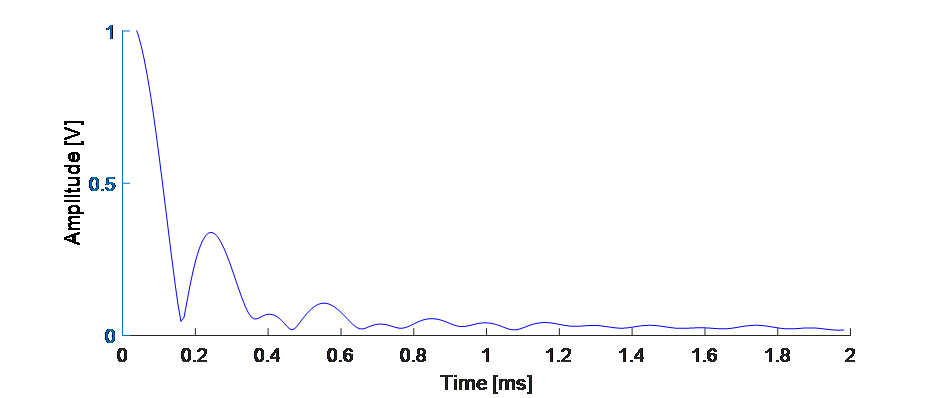}
	\caption{Real part of the deconvolved CEA response from a rubber sample, acquired with a chirp RF pulse of 32~kHz sweep range over 2~ms.}
	\label{fig:MR-Signal}
\end{figure}

\FloatBarrier
Fig.~\ref{fig:MR-Signal} shows a deconvolved MR signal response from the rubber phantom. 64~kHz chirp pulse was swept over 2~ms and the acquired signal was processed as described in~\cite{SWIFT}.
	
 Fig.~\ref{fig:4-line-leakage-eps-converted-to.pdf} shows the MRI signal after leakage subtraction from a rubber phantom with three tubes using fully-automatic controllable decoupling design. Two of the tubes contain CuSO4 with different densities.

 \begin{figure}[!ht]
	\centering
 	\includegraphics[width=6.8cm]{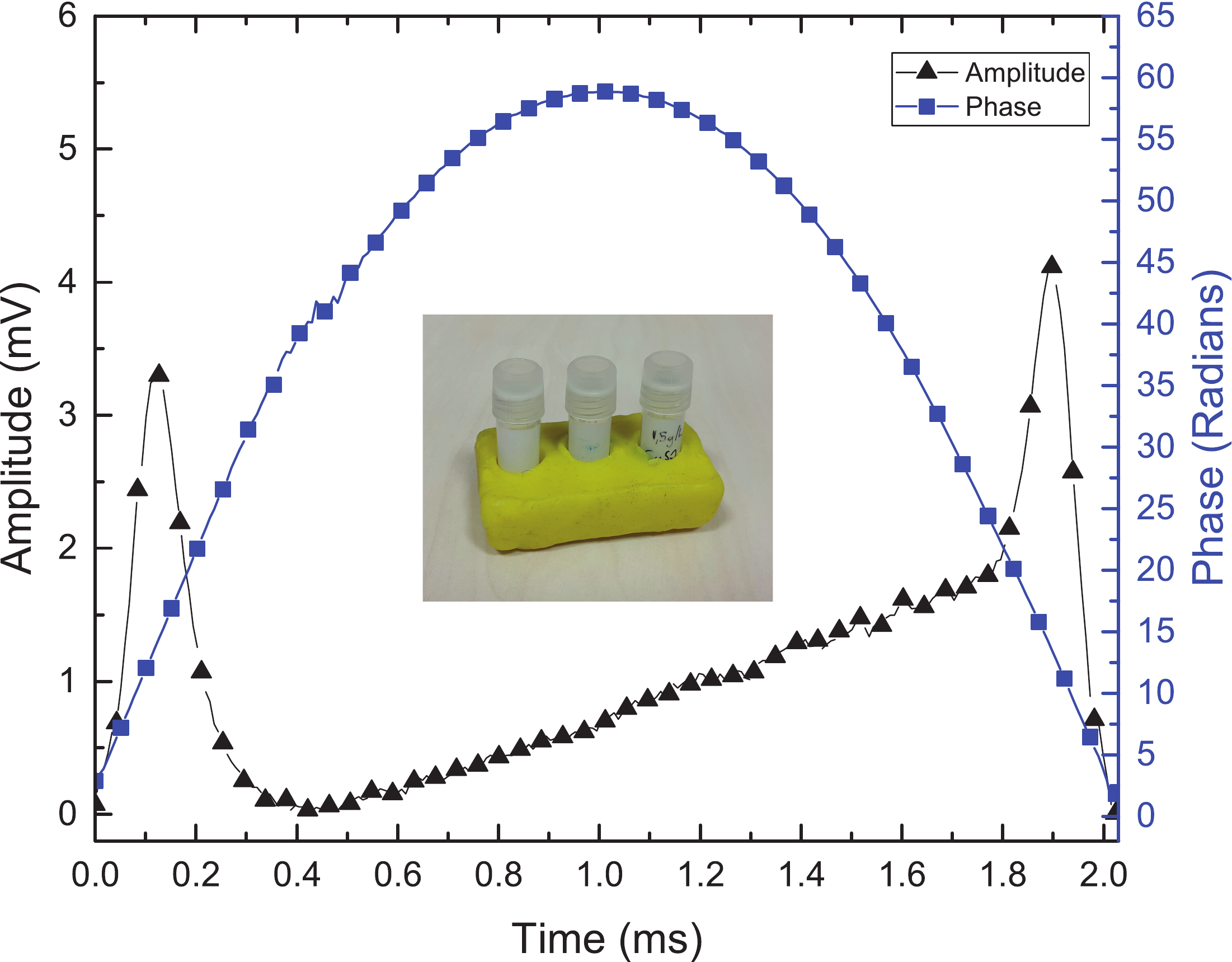}
	\caption{MR signal after leakage subtraction using fully-automatic controllable design.}
 	\label{fig:4-line-leakage}
 \end{figure}

\section{Conclusion}

We presented a new method to decouple transmit and receive coils for concurrent excitation and acquisition in MRI. The proof-of-concept demonstration of using automatically controllable decoupling circuits for CEA and getting MRI signal from tissues with very short $ T_2 $ time is done in this study. We show that the RF coupled signal in the receiver coil can be reduced more than 100~dB using our proposed method. This method consist of geometrical decoupling and automatically controlled decoupling design. We implemented the automatically controlled decoupling method using two designs: Semi-automatic and fully-automatic controllable decoupling designs. In the proposed method, cancellation circuit can automatically tune itself using a genetic algorithm optimization and provides favorable decoupling for CEA purposes. We demonstrated more than 100~dB decoupling using both designs and run MRI experiments using a clinical 3 Tesla MRI system with parallel transmit array to get MRI signal of a rubber phantom.

\section{Discussion}

\begin{figure}[!ht]
	\centering
	\includegraphics[width=8.5cm]{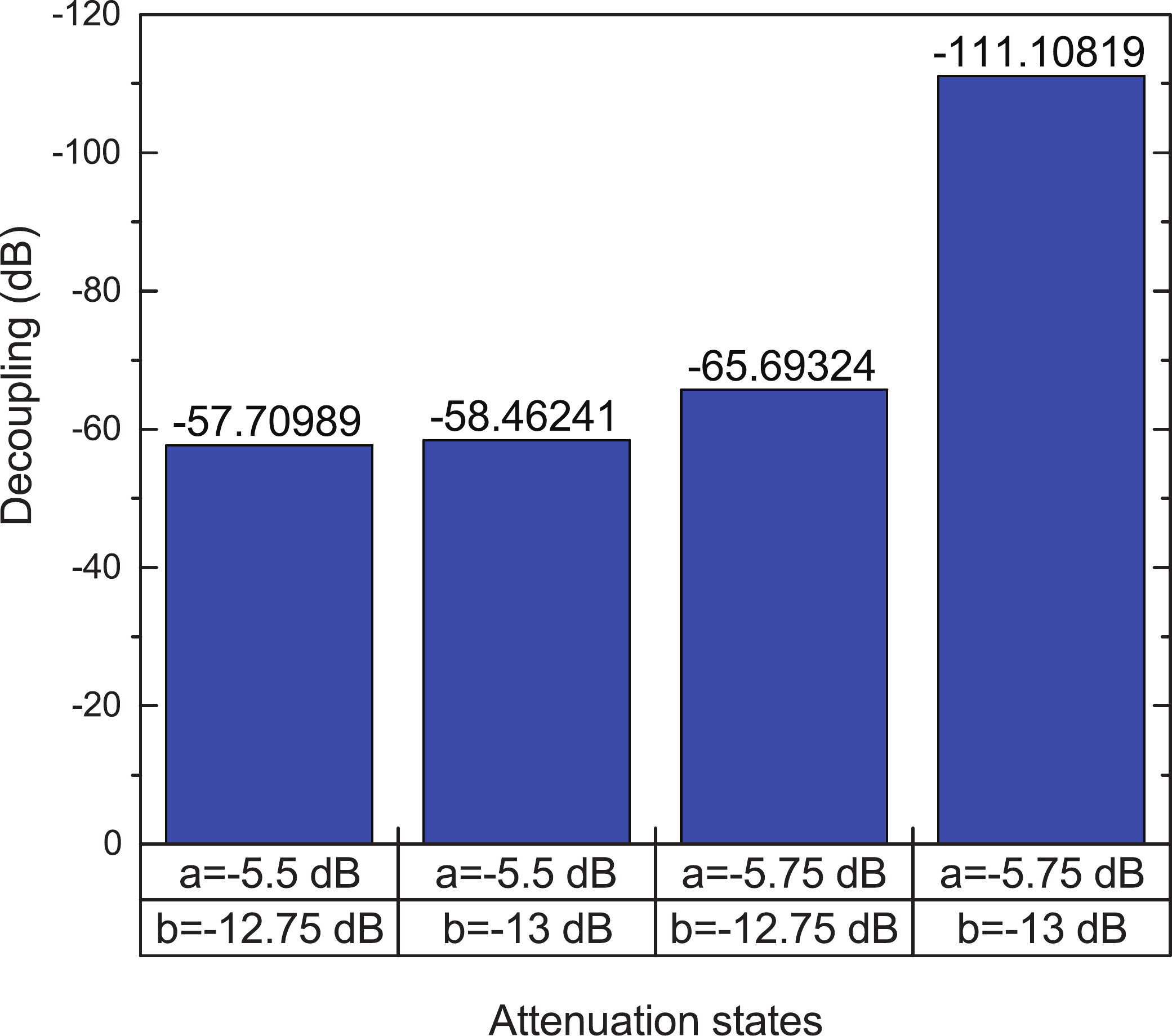}
	\caption{Amount of decoupling for the four nearest possible attenuation states to the desired circuit state. a=-5.7494~dB and b=-13~dB are the exact required attenuation amounts.}
	\label{fig:Descreet}
\end{figure}

Our design which is basically an analog cancellation circuit, while an important step forward, leaves several important new possibilities and future research. As we discussed, we need at least four lines consisting of a phase delay and an attenuator. In the mathematical approach that we discussed, the fixed phase delays are just created by the phase shifters. We have neglected the effect of the PCB tracks and other parts of the circuit such as attenuators and power dividers and combiners on the phase delay of each line. In the real case, each component adds an error to the phase delays and limits the maximum achievable decoupling. The attenuator IC that we have used has discreet available attenuation amounts. The phase delay caused by this IC varies from 5 to 20~degrees for different attenuation amounts. This affects the limitation on the maximum achievable decoupling which will be dependent on the phase delay of the coils caused by geometrical decoupling. In addition, discretization of ICs causes some drawbacks. For instance, for a special case, the attenuation of the two active lines should be 5.74 and 13~dB. But since the ICs cannot provide these exact attenuation amounts and the ICs have 0.25~dB step size, the decoupling provided by the circuit is limited. Fig.~\ref{fig:Descreet} shows amount of decoupling for the nearest possible attenuation states to the desired circuit state. Even a very small deviation from the best case decreases the amount of decoupling significantly.

Our current method is designed for single input single output (SISO) targets which is also practical for multiple input multiple output (MIMO). One can use the same design for transmit and receive array systems in CEA MRI applications. However, the key challenge is that for using N transmit chains N2 cancellation circuit is needed to achieve the same decoupling amount. The second point is that transmit coils should properly decoupled from each other by using capasitive decoupling techniques 

There is no feedback from MRI in our current prototype while CEA MRI is running. First, the circuit is tuned using network analyzer inside the MRI. Then, the network analyzer is disconnected from the circuit while we start the imaging without any feedback. This approach is suitable for static environments as well as dynamic environments with small changes over time. In order to make this design applicable for in vivo experiments and dynamic situations, a real-time feedback is required from the MRI scanner during the imaging. One can get the MRI signal before the reconstruction process, analyze it, and calculate the attenuation constants by solving the inverse problem of the system instead of the optimization algorithm. This process should be fast enough to be done between every two consecutive MRI signal pulses.


\section*{Acknowledgment}

The Scientific and Technological Research Council of Turkey (T{\"U}B\.{I}TAK) is gratefully acknowledged for funding through the project 114E186.

\ifCLASSOPTIONcaptionsoff
  \newpage
\fi

\bibliographystyle{IEEEtran}
\bibliography{whole}

\end{document}